\documentclass[onecolumn,secnumarabic,amssymb, amsmath, nofootinbib,tightenlines,
nobibnotes, aps, prl,epsfig]{revtex4}
\usepackage{graphicx}
\usepackage{dcolumn}
\usepackage{bm}
\begin{document}
\title{ Reduced cross section and gluon distribution in momentum-space approach }

\author{G.R. Boroun}%
 \email{boroun@razi.ac.ir }
 \affiliation{Department of Physics, Razi University, Kermanshah
67149, Iran}%
\author{Phuoc Ha }
\altaffiliation{pdha@towson.edu}
\affiliation{Department of Physics, Astronomy and Geosciences, Towson University, Towson, MD 21252}
\date{\today}
\begin{abstract}
We present a calculation of the reduced cross section in
momentum-space approach utilizing the Block-Durand-Ha (BDH)
parameterization of the proton structure function $F_{2}(x,Q^2)$
and the leading-order (LO) longitudinal structure function
$F_{L}(x,Q^2)$,  proposed by Boroun and Ha [G.R. Boroun and P.Ha,
Phys. Rev. D {\bf 109} (2024) 094037] using Laplace transform
techniques. Our results are compared with the HERA data and
extended to the Large Hadron electron Collider (LHeC) domain. We
also examine the ratio $F_{L2}(x, Q^2)=F_{L}(x, Q^2)/F_{2}(x,
Q^2)$ obtained from our work, comparing it with both the H1 data
and the color dipole (CDP) bounds. We find that our results for
the reduced cross section and the ratio $F_{L2}(x, Q^2)$ agree
with the H1 data. Finally, our evaluation of the gluon
distribution functions $G(x,Q^2)$ in momentum-space approach shows
very good concordance with the NNPDF3.0LO gluon structure
functions for moderate $Q^2$ in the range $10^{-5}{\leq}x{\leq}1$.

\end{abstract}
\maketitle
\section{Introduction}

One of the ways to understand and explore the dynamics of strong
interactions and test Quantum Chromodynamics (QCD) is through
measurements of the inclusive deep inelastic lepton-nucleon
scattering (DIS) cross-section \cite{Ref1}. The HERA accelerator,
which went through different phases known as HERA I  and HERA II,
operated from 1992 to 2007 at DESY in Hamburg. During this time,
electrons or positrons collided with protons at a center-of-mass
energy $\sqrt{s}{=}320~\mathrm{GeV}$. Specifically, one storage
ring accelerated electrons to energies of $27.6~\mathrm{GeV}$,
while the other accelerated protons to energies of
$920~\mathrm{GeV}$ in the opposite direction. HERA played a
crucial role in studying proton structure and quark properties,
laying the groundwork for research at the Large Hadron Collider
(LHC) at CERN. HERA kinematics cover the values of Bjorken-$x$ in
the interval $10^{-5}{\lesssim}x{<}1$ and $Q^2$, the squared
four-momentum transfer between lepton and nucleon, in the interval
$0.1~\mathrm{GeV}^2{\lesssim}Q^2{\lesssim}1000~\mathrm{GeV}^2$
\cite{Ref2}. These measurements can be performed with much
increased precision and extended to much lower values of $x$ and
high $Q^2$ in next ep colliders. These new colliders under design
are the Large Hadron electron Collider \cite{Ref3} with
center-of-mass energy $\sqrt{s}{\simeq}1.3~\mathrm{TeV}$ and the
Future Circular Collider electron-hadron (FCC-eh) \cite{Ref3,
Ref4} with $\sqrt{s}{=}3.5~\mathrm{TeV}$. The center-of-mass
energy at the LHeC is about 4 times of the center-of-mass energy
range of ep collisions at HERA and the kinematic range in the
$(x,Q^2)$ plane in neutral-current (NC) extends below
$x{\approx}10^{-6}$ and up to $Q{\simeq}1~\mathrm{TeV}$ and will
be extended down to $x{\approx}10^{-7}$  at the FCC-eh program
\cite{Ref3, Ref4, Ref5, Ref6, Ref7, Ref8}.\\

The measurements at HERA for the longitudinal structure function
have been performed with the extrapolation and derivative methods
at large and low $Q^2$ values \cite{Ref1}. At HERA, the
longitudinal structure function can be extracted from the
inclusive cross section only in the region of large inelasticity
with $y=Q^2/sx$, where $y$ is the inelasticity variable. Here $x$
and $Q^2$ are two independent kinematic variables and $s$ is the
center of mass energy squared. The reduced cross section, in the
inclusive DIS scattering, is defined in terms of the two proton
structure functions $F_{2}(x,Q^2)$ and $F_{L}(x,Q^2)$ as
\begin{eqnarray}
\sigma_{r}= F_{2}(x,Q^2)-\frac{y^2}{1+(1-y)^2}F_{L}(x,Q^2)
= F_{2}(x,Q^2)\bigg{[}1-\frac{y^2}{1+(1-y)^2}F_{L2}(x,Q^2)\bigg{]},
\end{eqnarray}
where $F_{L2}=F_{L}/F_{2}$. The ratio $F_{L2}$ can be defined into
the cross section ratio $R$ by the following form
\begin{eqnarray}
F_{L2}(x,Q^2)=\frac{F_{L}(x,Q^2)}{F_{2}(x,Q^2)}=\frac{R(x,Q^2)}{1+R(x,Q^2)},
\end{eqnarray}
where $R$ is related to the cross sections $\sigma_{T}$ and
$\sigma_{L}$ for absorption of transversely or longitudinally
polarised virtual photons as
\begin{eqnarray}
R(x,Q^2)=\frac{\sigma_{L}^{\gamma^{*}p}(x,Q^2)}{\sigma_{T}^{\gamma^{*}p}(x,Q^2)}.
\end{eqnarray}

In this paper, we present a calculation of the reduced cross
section in momentum-space approach using the Block-Durand-Ha (BDH)
parameterization\footnote{The BDH parameterization provides a
better fit to experimental data, particularly at low values of the
Bjorken variable $x$. This improved fit is crucial for accurately
describing the behavior of the proton structure function in
regions where data is sparse. Additionally, the BDH
parameterization aligns well with theoretical predictions, such as
the Froissart bound, which describes the asymptotic behavior of
hadron-hadron cross sections. By avoiding the need for a specific
factorization scheme, the BDH parameterization simplifies the
theoretical calculations involved in deep inelastic scattering
processes, making them more efficient and accessible.} of the
proton structure function $F_{2}(x,Q^2)$ \cite{Ref25} and the LO
longitudinal structure function $F_{L}(x,Q^2)$, proposed by Boroun
and Ha (BH) \cite{Ref26} using Laplace transform techniques
\cite{Ref27, Ref28, Ref29, Ref30}. We then compare our results of
the reduced cross section in the momentum-space approach with the
H1 data \cite{Ref31} and extend the results to the LHeC domain
\cite{Ref3}. Additionally, we investigate the ratio $F_{L2}(x,
Q^2)$ obtained from our work, comparing it with both the H1 data
and the color dipole picture (CDP) bounds\footnote{For further
discussion please refer to Appendix A.}. We also evaluate the
gluon distribution function in the momentum-space approach, and
compare our results with those in set NNPDF3.0 \footnote{The
NNPDF3.0 is the first set of parton distribution functions owing
to HERA, ATLAS, CMS and LHCb data based on LO, NLO and NNLO QCD
theory.} of the NNPDF Collaboration
of Ball {\em et. al} \cite{Ref32}.\\

\section{Reduced cross section in momentum-space approach}

The authors in Ref.\cite{Ref25} have obtained the BDH
parameterization of the structure function $F_{2}(x,Q^2)$, from a
combined fit to HERA data. The explicit expression takes
the following form
\begin{eqnarray}
F^{\mathrm{BDH}}_{ 2}(x,Q^{2})=
D(Q^{2})(1-x)^{n}\sum_{m=0}^{2}A_{m}(Q^{2})L^{m},
\end{eqnarray}
with
\begin{eqnarray}
D(Q^{2})&=&\frac{Q^{2}(Q^{2}+{\lambda}M^{2})}{(Q^{2}+M^{2})^2},
~ A_{0}=a_{00}+a_{01}L_{2},
 ~A_{i}(Q^{2})=\sum_{k=0}^{2}a_{ik}L_{2}^{k},~ (i=1,2),\nonumber\\
& &L=\ln(1/x)+L_{1},~ L_{1}={\ln}\frac{Q^{2}}{Q^{2}+\mu^{2}},~ L_{2}={\ln}\frac{Q^{2}+\mu^{2}}{\mu^{2}},\nonumber\\
\end{eqnarray}
 where the effective parameters are summarized in
Ref.\cite{Ref33} and are given in Table I.

In a recent paper \cite{Ref26}, using Laplace transform techniques
\cite{Ref27, Ref28, Ref29, Ref30},  we  have determined the
longitudinal structure function  $F_{L}(x,Q^2)$, at the
leading-order approximation in momentum-space approach\footnote{The momentum-space approach has two advantages:\\
1) It is no need to define a factorization scheme,\\
2) The approach in terms of physical structure functions has the
advantage of being more transparent in the parametrization of the
initial conditions of the evolution.}, from the proton structure
function and its derivative with respect to $\ln{Q^2}$ as
\begin{eqnarray}
F^{\mathrm{BH}}_{L}(x,Q^2)&=&4\int_{x}^{1}\frac{d{F}^{BDH}_{2}(z,Q^2)}{d{\ln}Q^2}(\frac{x}{z})^{3/2}\bigg{[}\cos{\bigg{(}}\frac{\sqrt{7}}{2}{\ln}\frac{z}{x}{\bigg{)}}-\frac{\sqrt{7}}{7}
\sin{\bigg{(}}\frac{\sqrt{7}}{2}{\ln}\frac{z}{x}{\bigg{)}}\bigg{]}\frac{dz}{z}-4C_{F}\frac{\alpha_{s}(Q^2)}{2\pi}\nonumber\\
&&{\times}\int_{x}^{1}F^{BDH}_{2}(z,Q^2)(\frac{x}{z})^{3/2}\bigg{[}(1.6817+2\psi(1))\cos{\bigg{(}}\frac{\sqrt{7}}{2}{\ln}\frac{z}{x}{\bigg{)}}
+(2.9542-2\frac{\sqrt{7}}{7}\psi(1))
\sin{\bigg{(}}\frac{\sqrt{7}}{2}{\ln}\frac{z}{x}{\bigg{)}}\bigg{]}\frac{dz}{z}\nonumber\\
&&+8C_{F}\frac{\alpha_{s}(Q^2)}{2\pi}
\sum_{k=1}^{\infty}\frac{k}{(k+1)^2-3(k+1)+4}\int_{x}^{1}F_{2}^{BDH}(z,Q^2)(\frac{x}{z})^{k+1}
\frac{dz}{z}.
\end{eqnarray}
As shown in the Appendix B,  the last term in Eq. (6)  can be
modified to improve the convergence substantially for increasing
numbers of terms in the series. Therefore, the momentum space
evolution of the longitudinal structure function
$F^{\mathrm{BH}}_{L}(x,Q^2)$ can then be written as
\begin{eqnarray}
F^{\mathrm{BH}}_{L}(x,Q^2)&=&4\int_{x}^{1}\frac{d{F}^{BDH}_{2}(z,Q^2)}{d{\ln}Q^2}(\frac{x}{z})^{3/2}\bigg{[}\cos{\bigg{(}}\frac{\sqrt{7}}{2}{\ln}\frac{z}{x}{\bigg{)}}-\frac{\sqrt{7}}{7}
\sin{\bigg{(}}\frac{\sqrt{7}}{2}{\ln}\frac{z}{x}{\bigg{)}}\bigg{]}\frac{dz}{z}-4C_{F}\frac{\alpha_{s}(Q^2)}{2\pi}\nonumber\\
&&{\times}\int_{x}^{1}F^{BDH}_{2}(z,Q^2)(\frac{x}{z})^{3/2}\bigg{[}(1.6817+2\psi(1))\cos{\bigg{(}}\frac{\sqrt{7}}{2}{\ln}\frac{z}{x}{\bigg{)}}
+(2.9542-2\frac{\sqrt{7}}{7}\psi(1))
\sin{\bigg{(}}\frac{\sqrt{7}}{2}{\ln}\frac{z}{x}{\bigg{)}}\bigg{]}\frac{dz}{z}\nonumber\\
&&+8C_{F}\frac{\alpha_{s}(Q^2)}{2\pi}
\bigg{[}\sum_{m=1}^{\infty}\bigg{(}\frac{2(m-4)}{m(m^2-3m+4)}
-\frac{2}{m^2} \bigg{)}\int_{x}^{1}F_{2}^{BDH}(z,Q^2)(\frac{x}{z})^{m}\frac{dz}{z}\nonumber\\
&&
+ \int_{x}^{1}F_{2}^{BDH}(z,Q^2)\bigg{(}{\rm Li}_ 2 (\frac{x}{z}) -\ln(1-\frac{x}{z})\bigg{)} \frac{dz}{z} \bigg{]}.
\end{eqnarray}

\begin{figure}[h]
\includegraphics[width=0.6\textwidth]{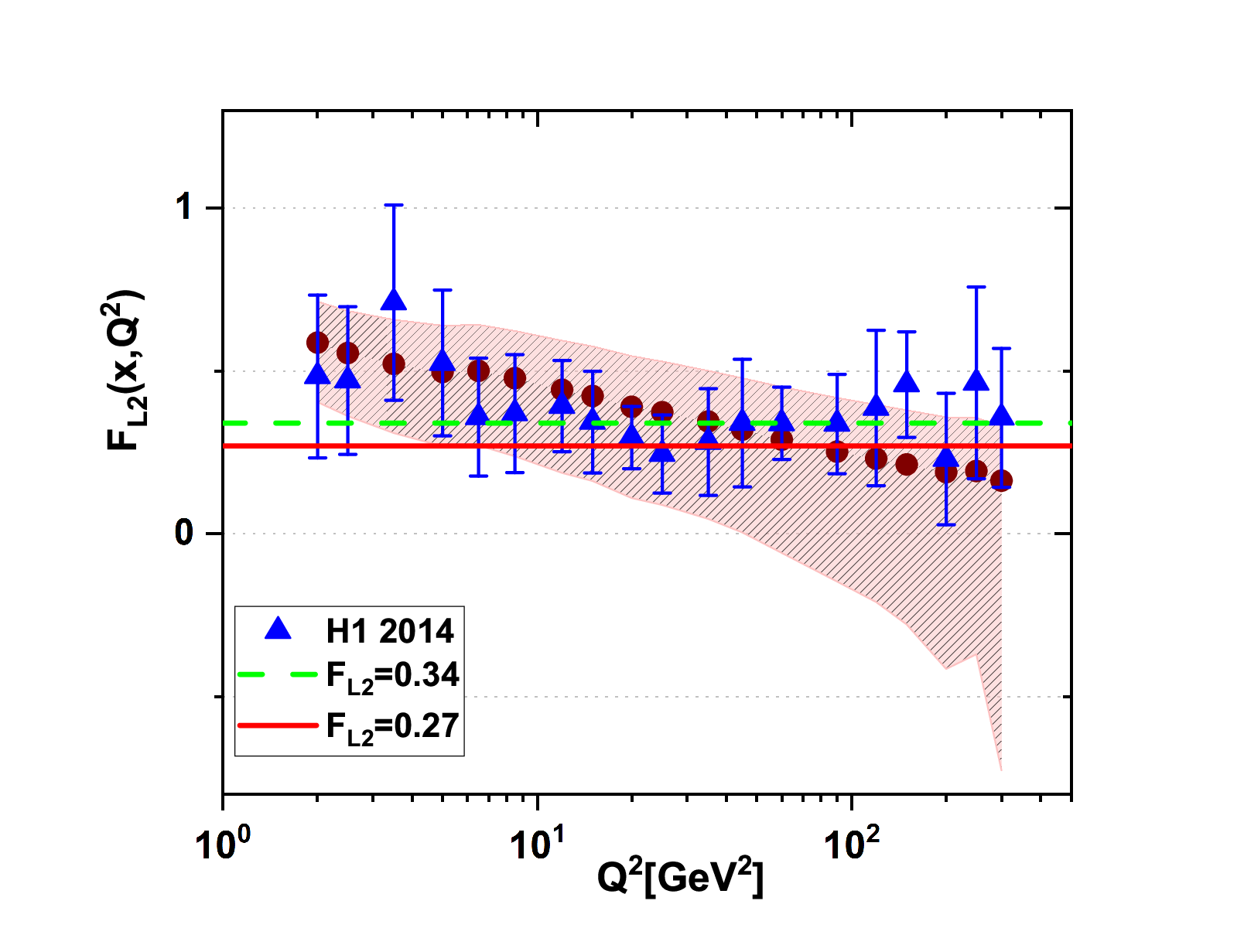}
\caption{The extracted ratio $F_{L2}$ ( shown as brown points)
from the parametrization methods
 is compared with the H1 data \cite{Ref31}. The total errors are included,
 and the
 dipole upper bounds are represented by
curves corresponding to $F_{L2}=0.27$ (solid red line) and
$F_{L2}=0.34$ (dashed green line). The error  bands (rose gold) of
the ratio $F_{L2}$ correspond to the uncertainty in the
parameterization of $F_{2}(x, Q^2)$ in \cite{Ref25}.}\label{Fig1}
\end{figure}
In Fig.1, we show the ratio of the structure functions based on
the $F_{2}$ and $F_{L}$ parametrizations in \cite{Ref25} and
\cite{Ref26}, respectively. The behavior of the ratio
$F_{L2}(x,Q^2)$ is compared with the H1 data \cite{Ref31} and the
CDP bounds. The H1 data are selected in the region
$2{\leq}Q^2{\leq}300~\mathrm{GeV}^2$ at the interval
$156{\leq}W{\leq}233~\mathrm{GeV}$ with the maximum value of the
longitudinal structure function\footnote{Please see Table 5 in
Ref.\cite{Ref31}}.
The values of the ratio of structure functions describe well the
H1 data  in the region $2{\leq}Q^2{\leq}300~\mathrm{GeV}^2$ and
they are in good agreement with the CDP bounds in the interval
$5{\leq}Q^2{\leq}300~\mathrm{GeV}^2$ as data on $F_{L2}(x,Q^2)$
confirm the standard dipole picture at these kinematic points. The
error bars of the ratio $F_{L2}$ (in H1 data and our method) are
determined by the following form:
 $$\Delta({F_{L2}})=F_{L2}\sqrt{(\Delta{F_{L}}/F_{L})^2+(\Delta{F_{2}}/F_{2})^2},$$
where in the H1 data, $\Delta{F_{L}}$ and  $\Delta{F_{2}}$ are
collected from the H1 experimental data \cite{Ref31}. In this
paper, the error bands are dependent on the uncertainties of
$F_{2}(x,Q^2)$ and $F_{L}(x,Q^2)$. The uncertainties in
$F_{2}(x,Q^2)$ are obtained from the parametrization coefficients
in the BDH model (Table I), while the uncertainties in
$F_{L}(x,Q^2)$ are obtained through Eq.(7), which  depends on the
uncertainties in $F_{2}(x,Q^2)$.\\
The values of the ratio of structure
functions are comparable with the H1 data and they are in good
agreement with the CDP bounds in the interval
$5{\leq}Q^2{\leq}300~\mathrm{GeV}^2$ as data on $F_{L2}(x,Q^2)$
confirm the standard dipole picture at these kinematic points. In
order to include the effect of production threshold for charm quark
 with $m_{c}=1.29^{+0.077}_{-0.053}~\mathrm{GeV}$ \cite{Ref31, Ref34}, the
rescaling variable $\chi$ is defined by the form
$\chi=x(1+4\frac{m_{c}^{2}}{Q^2})$ where reduced to the Bjorken
variable $x$ at high $Q^2$ \cite{Ref34}. The rescaling variable is
one of the ingredients used in the general-mass variable flavor
number schemes (GM-VFNS)\cite{Ref35}. Using $\alpha_s(M_Z^2) =
0.1166$ and the leading-order (LO) form of $\alpha_s(Q^2)$, the
QCD parameter $\Lambda$ for four active flavors has been extracted
[33], resulting in $\Lambda = 136.8$~MeV.\\
\begin{figure}[h]
\includegraphics[width=0.6\textwidth]{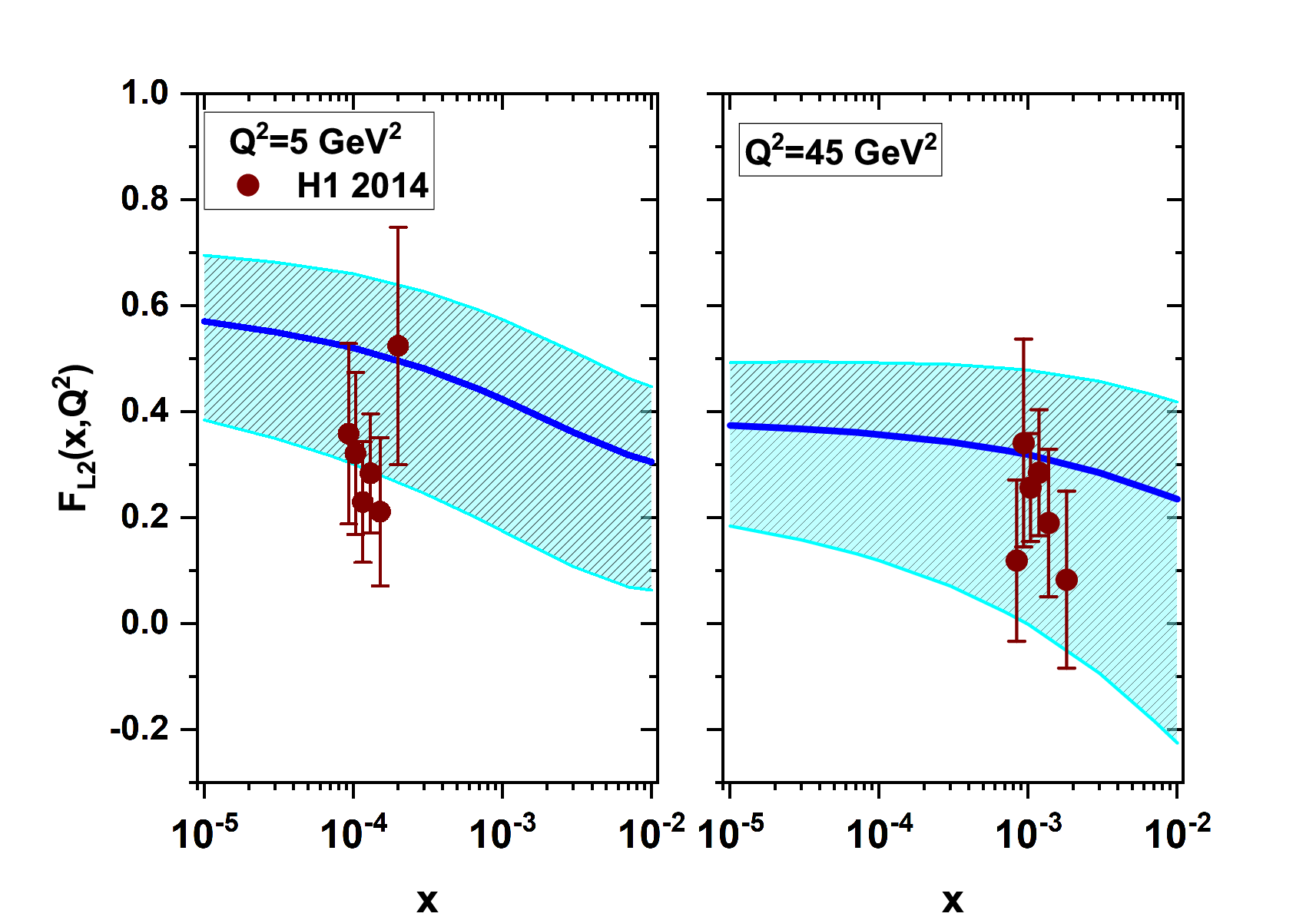}
\caption{The ratio of structure functions extracted  (blue solid curve) in comparison
with the H1 data \cite{Ref31} as accompanied with total errors.
The results are presented at $Q^2=5$ and $45~\mathrm{GeV}^2$ in a
wide range of $x$. The error (turquoise) bands correspond to the uncertainty
in the parameterization of $F_{2}(x, Q^2)$ in \cite{Ref25}.
}\label{Fig2}
\end{figure}

In Fig. 2, the results for the ratio of structure functions,
$F_{L2}(x,Q^2)$, at fixed values of $Q^2=5$ and
$45~\mathrm{GeV}^2$ in a wide range of $x$ are presented and
compared with the H1 data \cite{Ref31} as accompanied with total
errors. The error bands correspond to the uncertainty in the
parameterization of $F_{2}(x, Q^2)$ in \cite{Ref25}. As seen in
the figure, the results are comparable with the H1 data in a wide
range of $x$. The extracted ratio of structure functions is in
good agreement with the H1 data as accompanied with total
errors.\\
The extracted results for the longitudinal structure function in
momentum-space approach Ref.\cite{Ref26} are in line with data
from the H1 Collaboration and other results using Mellin transform
method \cite{Ref33}. In the following, the ratio
$F_{L2}(x,Q^2)=F_{L}^{\mathrm{BH}}(x,Q^2)/F_{2}^{\mathrm{BDH}}(x,Q^2)$
is parametrized using the BDH parametrization of the proton
structure function $F_{2}^{\mathrm{BDH}}(x,Q^2)$ (i.e., Eq.(4)),
and the reduced cross section is parametrized by
\begin{eqnarray}
\sigma_{r}=F_{2}^{\mathrm{BDH}}(x,Q^2)\bigg{[}1-\frac{y^2}{1+(1-y)^2}F_{L2}(x,Q^2)\bigg{]}.
\end{eqnarray}
The calculation of the ratio of structure functions facilitates the accurate determination
 of the reduced cross section $\sigma_{r}$ (i.e.,
Eq.(8)). The results of the reduced cross section $\sigma_{r}$ are
depicted in Fig.3 as the center-of-mass energy extended to the
LHeC study group \cite{Ref4}. A comparison with the H1 data
\cite{Ref1} is done as accompanied with total errors at moderate
$x$ and extended to the very low $x$ due to the LHeC region with
$y{\leq}1$. These results for the reduced cross section reflect
the large extension of kinematic range towards low $x$ and high
$Q^2$ available at the LHeC, as
compared to HERA.\\
\begin{figure}[h]
\includegraphics[width=0.6\textwidth]{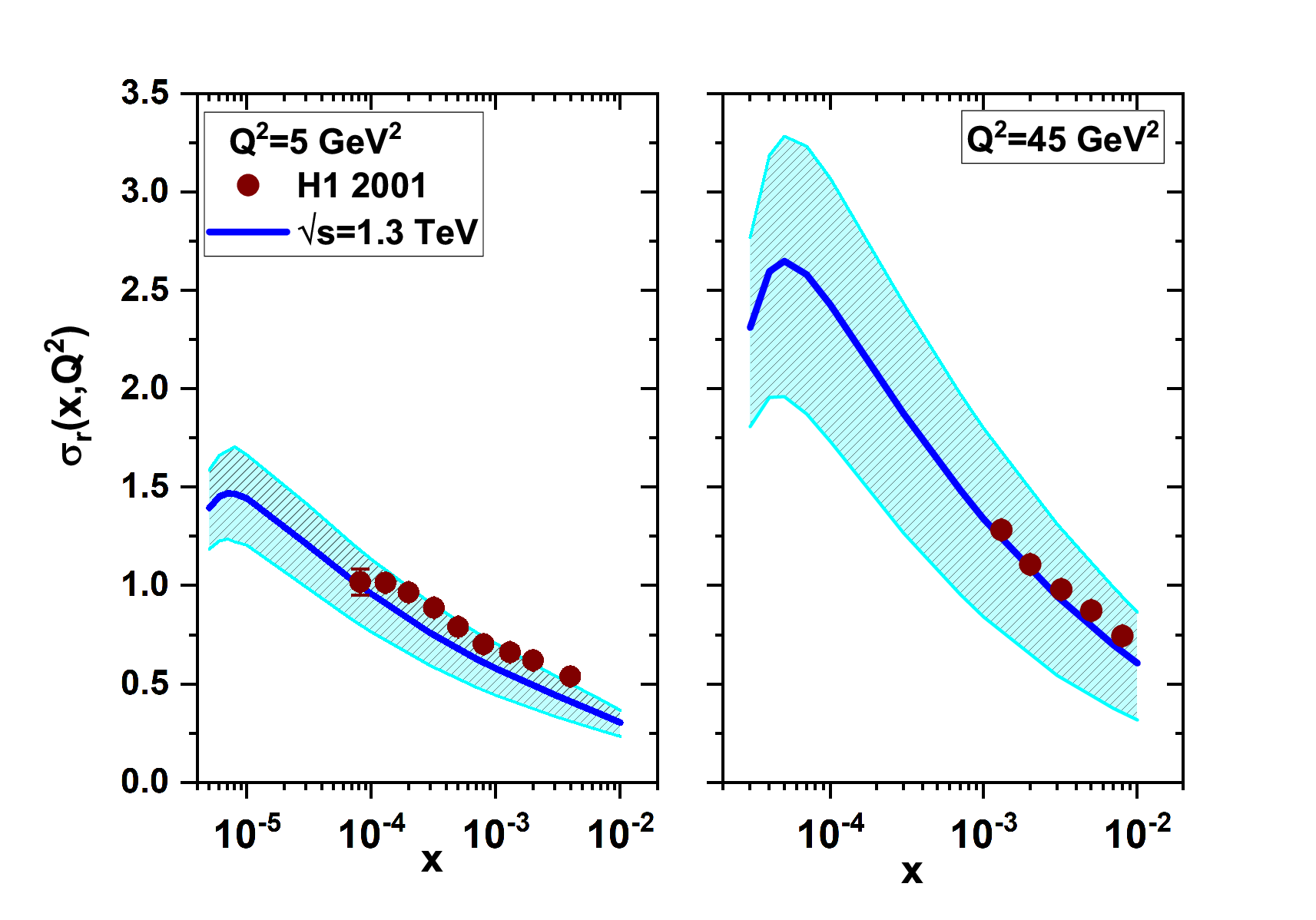}
\caption{The reduced cross section extracted  (blue solid curve) in comparison with
the H1 data \cite{Ref1} as accompanied with total errors. The
results are presented and extended to  the center-of-mass energy
of the LHeC, $\sqrt{s}=1.3~\mathrm{TeV}$ with $y<1$ at $Q^2=5$ and
$45~\mathrm{GeV}^2$. The error (turquoise) bands correspond to the uncertainty
in the parameterization of $F_{2}(x, Q^2)$ in \cite{Ref25}
}\label{Fig3}
\end{figure}

\section{Gluon distribution in momentum-space approach}

The gluonic density, in high-energy scattering processes, exhibits
a crucial phenomenon at the small-$x$ region and plays a vital
role in estimating backgrounds. At low values of $x$, the
structure functions $F_{2}(x,Q^{2})$ and  $F_{L}(x,Q^{2})$ are
defined solely via the singlet quark $x\Sigma(x,Q^{2})$ and gluon
distribution $x{g}(x,Q^{2})$ as
\begin{eqnarray}
F_{k}(x,Q^{2})=<e^{2}>\bigg{[}B_{k,s}(x, \alpha_{s}(Q^2)
){\otimes}x\Sigma(x,Q^{2})+B_{k,g}(x,
\alpha_{s}(Q^2)){\otimes}xg(x,Q^{2}) \bigg{]},~~~k=2,L \nonumber
\end{eqnarray}
where $<e^{2}>$ is the average  charge squared for  the number of
effective flavours and $\alpha_{s}(Q^2)$ is the running coupling.
The quantities $B_{k,a}(x)$ are the known Wilson coefficient
functions and the parton densities fulfil the renormalization
group evolution equations\footnote{Here the non-singlet densities
become negligibly small in comparison with the singlet densities.
The symbol $\otimes$ indicates convolution over the variable $x$
by the usual form, $f(x){\otimes}g(x)=\int_{x}^{1}
\frac{dz}{z}f(z,\alpha_{s})g(x/z)$.}.

The gluon density in the momentum-space approach contributes to
the DIS structure functions $F_{2}$ and $F_{L}$ as defined in
\cite{Ref36} by the following formula:
\begin{eqnarray}
g(x,Q^2)=\int_{x}^{1}\frac{dz}{z}\delta(1-z)\bigg{\{}\eta\bigg{[}\frac{x}{z}\frac{d}{d\frac{x}{z}}-2\bigg{]}
\frac{F_{2}(\frac{x}{z},
Q^2)}{\frac{x}{z}}+\zeta\bigg{[}\frac{x^2}{z^2}\frac{d^2}{d(\frac{x}{z})^2}-2\frac{x}{z}\frac{d}{d\frac{x}{z}}+2\bigg{]}
\frac{F_{L}(\frac{x}{z},
Q^2)}{\frac{x}{z}\frac{\alpha_{s}(Q^2)}{2\pi}}\bigg{\}},
\end{eqnarray}
where $\eta={C_{F}}/({4T_{R}n_{f}<e^2>})$ and
$\zeta={1}/({8T_{R}n_{f}<e^2>})$ with the color factors
$T_{R}=1/2$ and $C_{F}=4/3$ associated with the color group
SU(3).

After successive differentiations of the brackets in Eq. (9) with
respect to $\frac{x}{z}$ and some rearranging, using the identity
$\frac{x}{z}=y$, we find
\begin{eqnarray}
g(x,Q^2)&=&\int_{x}^{1}\frac{dy}{y}\delta(1-\frac{x}{y})\bigg{\{}\eta\bigg{[}\frac{{\partial}F_{2}(y,Q^2)}{{\partial}y}-3\frac{F_{2}(y,
Q^2)}{y}\bigg{]} \nonumber \\
&&+\zeta\frac{2\pi}{\alpha_{s}(Q^2)}\bigg{[}{y}\frac{{\partial}^2F_{L}(y,
Q^2)}{{\partial}^2y}-4\frac{{\partial}F_{L}(y,
Q^2)}{{\partial}y}+\frac{6}{y}F_{L}(y, Q^2)\bigg{]} \bigg{\}}.
\end{eqnarray}
Using the delta function property, we find the the momentum space
evolution of the gluon distribution in terms of the structure
functions as
\begin{eqnarray}
G(x,Q^2)=\eta\bigg{[}x\frac{{\partial}F_{2}(x,Q^2)}{{\partial}x}-3{F_{2}(x,
Q^2)}\bigg{]}
+\zeta\frac{2\pi}{\alpha_{s}(Q^2)}\bigg{[}{x^2}\frac{{\partial}^2F_{L}(x,
Q^2)}{{\partial}^2x}-4x\frac{{\partial}F_{L}(x,
Q^2)}{{\partial}x}+{6}F_{L}(x, Q^2)\bigg{]},
\end{eqnarray}
where $G(x,Q^2)=xg(x,Q^2)$. This is a simple form of the gluon
distribution, expressed through the parametrization of the proton
structure function (i.e., Eq. (4)) and the longitudinal structure
function (i.e., Eq. (7)) in momentum-space approach at low values
of $x$. We note that the method of Lappi et al. \cite{Ref36}, on
which our work is based, (1) neglects non-singlet contributions
and (2) is a high-$Q^2$ approximation. The results for the gluon
distribution function  (i.e. Eq.(11)) in the momentum-space
approach are presented in Figs.4-5. These results are included
with and without the rescaling variable, including the charm quark
mass.\\
\begin{figure}[h]
\includegraphics[width=0.8\textwidth]{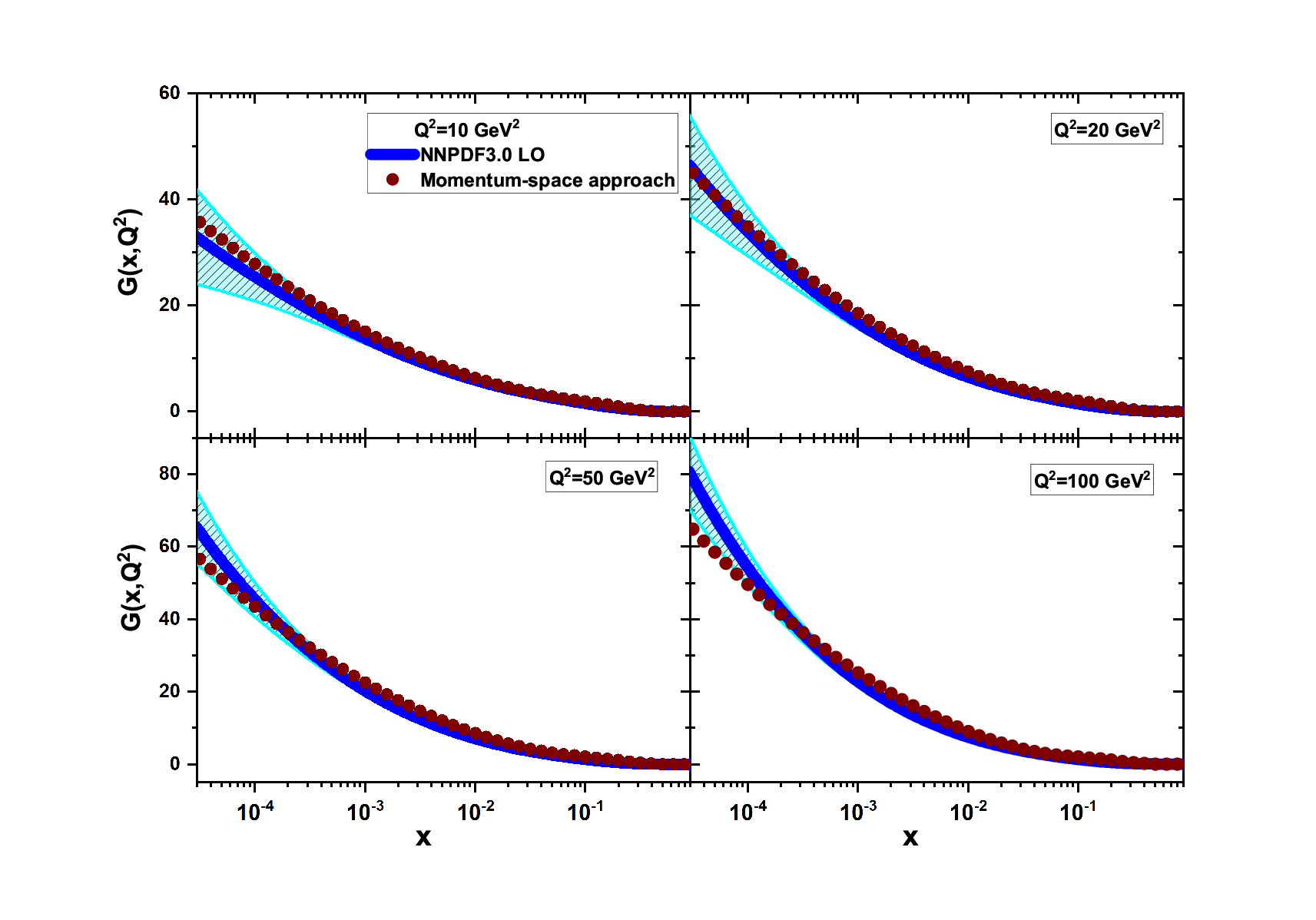}
\caption{The gluon distribution function $G(x,Q^2)$ (brown points)
with considering the rescaling variable of $x$ in the
momentum-space approach extracted and compared with the NNPDF3.0LO
\cite{Ref32} as accompanied with total errors (turquoise bands)
for $Q^2=10, 20, 50$, and $100~\mathrm{GeV}^2$ in a wide range of
$x$.}\label{Fig4}
\end{figure}
\begin{figure}[h]
\includegraphics[width=0.7\textwidth]{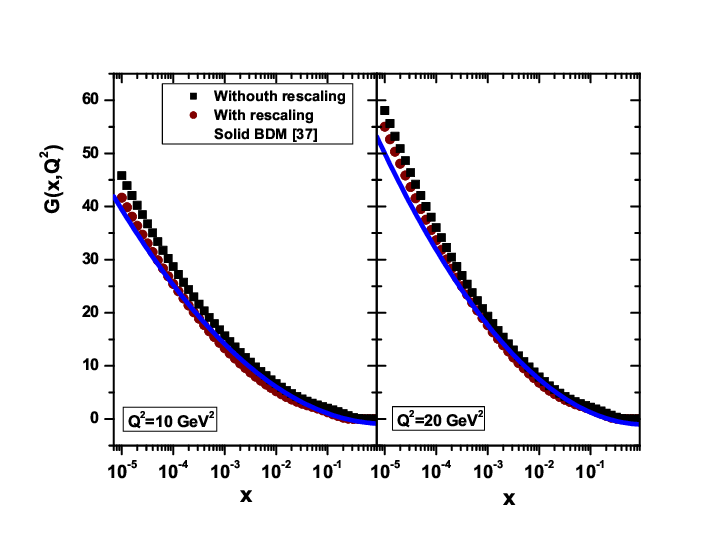}
\caption{The gluon distribution function $G(x,Q^2)$ with (brown
points) and without (black points) rescaling variable in the
momentum-space approach extracted and compared with the BDM result
\cite{Ref37} (blue curve) for $Q^2=10$,  and $20~\mathrm{GeV}^2$
in a wide range of $x$.}\label{Fig5}
\end{figure}
As can be seen in Fig.4, the results are comparable with the
NNPDF3.0LO for $Q^2=10, 20, 50$, and $100~\mathrm{GeV}^2$ in a
wide range of $x$ with the rescaling variable of $x$. Notably, the
values of the gluon distribution function increase as $x$
decreases, a trend that is in harmony with the expectations of
pQCD. The results of $G(x,Q^2)$ based on the momentum-space
approach show very good agreement with the NNPDF3.0LO gluon
structure function for moderate $Q^2$ in the range
$10^{-5}{\leq}x{\leq}1$. The comparison between the results (i.e.,
momentum-space approach and NNPDF3.0LO) with the rescaling
variable at $Q^2=10, 20$ and $50~\mathrm{GeV}^2$ in a wide range
of $x$ is excellent. In Fig.5, we compared the gluon distribution
results with and without considering the rescaling variable of $x$
at $Q^2=10$ and $20~\mathrm{GeV}^2$ with an analytical solution
using a Froissart bounded structure function $F_{2}(x,Q^2)$ in
Ref.\cite{Ref37} by M. M. Block, L. Durand and Douglas W. McKay
(BDM). We observe that the results with the rescaling variable is
comparable with those using the Froissart-bound type
parametrization of the proton
structure function.\\

\section{Conclusions}
The manuscript focuses on three interconnected topics: the study
of the ratio $F_{L2}(x, Q^2) = F_L(x, Q^2)/F_2(x, Q^2)$, the
calculation of the reduced cross section in momentum space using
the Block-Durand-Ha (BDH) parameterization of the proton structure
function $F_2(x, Q^2)$ and the leading-order (LO) longitudinal
structure function $F_L(x, Q^2)$, and the evaluation of the gluon
distribution function $G(x, Q^2)$.\\

We have developed a method for the analytical solution of the
longitudinal structure function based on the proton structure
function at low-$x$ in the momentum-space approach. The momentum
space evolution of the longitudinal structure function relies on
the Froissart-bounded parametrization of the DIS structure
function $F_{2}(x,Q^2)$ within the leading-order approximation.\\
The extraction procedure has been elaborated for an analysis of
the ratio of structure functions, $F_{L2}(x,Q^2)=
\frac{F_{L}(x,Q^2)}{F_{2}(x,Q^2)}$ in the kinematical region of
the H1 collaboration data. The momentum space evolution of
 the ratio of structure functions is entirely determined by the
 effective parameters of the BDH parametrization and compared with
 the CDP bounds. The study further examines the ratio $F_{L2}(x, Q^2)$, comparing it to
H1 data and bounds from the color dipole model, finding strong
agreement that supports the proposed methods.\\

 We have applied the developed method to extract the
reduced cross section in the momentum-space approach within the
kinematical conditions corresponding to that available at the HERA
collider. The results are validated against HERA data and extended
to the Large Hadron Electron Collider (LHeC) domain. We have
performed an analysis of the $x$-evolution of the extracted of the
gluon distribution function in momentum-space approach.\\
 It has
been demonstrated that the gluon distribution function $G(x, Q^2)$
is evaluated in momentum-space approach and compared to NNPDF3.0LO
gluon structure functions, demonstrating excellent agreement for
moderate $Q^2$ values within the range $10^{-5} \leq x \leq 1$.
Such an analysis can be accomplished by employing the charm quark
mass in the rescaling variable (which is directly related to the
gluon density in the photo-gluon fusion reactions) as a useful
tool for estimations of the gluon distributions in comparison with
other results.\\

In summary, our calculation of the reduced cross section in
momentum-space approach employs the Block-Durand-Ha
parameterization for the proton structure function  $F_{2}(x,Q^2)$
 and the LO longitudinal structure function
$F_{L}(x,Q^2)$, as proposed by Boroun and Ha, utilizing Laplace transform techniques.
We have benchmarked our reduced cross section results against the H1 data
and extrapolated them into the LHeC domain. Furthermore, the ratio $F_{L2}(x, Q^2)$ derived
from our analysis is compared with both the H1 data and the CDP bounds, showing consistency.
Lastly, our evaluation of the gluon distribution functions $G(x,Q^2)$
 in momentum-space approach corroborates the NNPDF3.0LO  gluon distribution functions for moderate $Q^2$
 values within the range $10^{-5}{\leq}x{\leq}1$. The results not
 only are comparable with the NNPDF3.0LO  gluon distribution
 functions but also serve the Froissart-bound type parametrization of the proton
structure function.\\

\section{ACKNOWLEDGMENTS}

Authors would like to thank Professor Loyal Durand for useful
comments and invaluable support.

\section{Appendix A}

\section{Color Dipole Model}

In the color dipole model (CDM), the virtual photon $\gamma^{*}$
exchanged between the electron and proton currents with virtuality
$Q^2$, split into a quark-antiquark pair (a dipole) which then
interacts with the target proton via gluon exchanges \cite{Ref9}.
The dipole picture for DIS is used to describe the data at low and
moderate values of $Q^2$, as shown in the various applications of
the dipole model in Refs.~\cite{Ref10, Ref11, Ref12, Ref13, Ref14,
Ref15, Ref16}. The total $\gamma^{*}p$ cross section is given by
\begin{eqnarray}
\sigma_{L,T}^{\gamma^{*}p}(x,Q^{2})=\sum_{f}\int d^{2}\mathbf{r}
\int_{0}^{1} dz
|\Psi_{L,T}(\mathbf{r},z;Q^{2})|^{2}\sigma_{\mathrm{dip}}({x},\mathbf{r}),
\end{eqnarray}
where the sum over quark flavours {\it f} is performed. The quark
and antiquark in this dipole carry a fraction $z$ and $1-z$ of the
photon longitudinal momentum respectively, and  the transverse
size between the quark and antiquark is given by the vector
$\mathbf{r}$. Here $\Psi_{L,T}(\mathbf{r},z;Q^2)$ are the
appropriate spin averaged light-cone wave functions of the photon,
which give the probability for the occurrence of a
$(q\overline{q})$ fluctuation of transverse size with respect to
the photon polarization.\\

The measured structure functions $F_{2}$ and $F_{L}$ are related
to the dipole cross section $\sigma_{\mathrm{dip}}$ by
\begin{eqnarray}
F_{2}(x,Q^{2})=\frac{Q^2}{4\pi^2\alpha_{\mathrm{em}}}(1-x)\sum_{f}\int
d^{2}\mathbf{r} \int_{0}^{1} dz
\bigg{[}|\Psi_{T}(\mathbf{r},z;Q^{2})|^{2}+|\Psi_{L}(\mathbf{r},z;Q^{2})|^{2}\bigg{]}\sigma_{\mathrm{dip}}({x},\mathbf{r}),
\end{eqnarray}
and
\begin{eqnarray}
F_{L}(x,Q^{2})=\frac{Q^2}{4\pi^2\alpha_{\mathrm{em}}}(1-x)\sum_{f}\int
d^{2}\mathbf{r} \int_{0}^{1} dz
|\Psi_{L}(\mathbf{r},z;Q^{2})|^{2}\sigma_{\mathrm{dip}}({x},\mathbf{r}).
\end{eqnarray}
The ratio of structure functions is defined by the following form
\begin{eqnarray}
F_{L2}(x,Q^{2})=\frac{\sum_{f}\int d^{2}\mathbf{r} \int_{0}^{1} dz
|\Psi_{L}(\mathbf{r},z;Q^{2})|^{2}\sigma_{\mathrm{dip}}({x},\mathbf{r})}{\sum_{f}\int
d^{2}\mathbf{r} \int_{0}^{1} dz
\bigg{[}|\Psi_{T}(\mathbf{r},z;Q^{2})|^{2}+|\Psi_{L}(\mathbf{r},z;Q^{2})|^{2}\bigg{]}\sigma_{\mathrm{dip}}({x},\mathbf{r})}.
\end{eqnarray}

In Refs.\cite{Ref14, Ref15, Ref17}, authors show that at large
$Q^2$, the ratio of photo absorption cross sections
 is determined by a parameter $\rho$ that describes the dissociation
 of photons into $q\bar{q}$ pairs, $\gamma^*_{L,T}\rightarrow q\bar{q}$, with
\begin{eqnarray}
R=\frac{1}{2\rho},
\end{eqnarray}
where the factor 2 originates from the difference in the photon
wave functions. Indeed, the $\rho$ parameter describes the ratio
of the average transverse momenta
$\rho=\frac{<\overrightarrow{k}^{2}_{\bot}>_{L}}{<\overrightarrow{k}^{2}_{\bot}>_{T}}$,
or it can  be related to the ratio of the effective transverse
sizes of the $(q\overline{q})^{J=1}_{L,T}$ states as
$\frac{<\overrightarrow{r}^{2}_{\bot}>_{L}}{<\overrightarrow{r}^{2}_{\bot}>_{T}}=\frac{1}{\rho}$.
For the parameter $\rho=\frac{4}{3}$ in Ref.\cite{Ref15}
\footnote{For further discussion, readers can refer to the papers
by M. Kuroda and D. Schildknecht: "Phys. Rev. D 85  (2012) 094001"
and
 "J. Mod. Phys. A 31 (2016)
1650157". }, one can find that $R=\frac{3}{8}=0.375$ and the ratio
of structure functions is $F_{L2}(x,Q^2)=\frac{3}{11}=0.273$. For
the specific value $\rho=1$ (i.e., helicity independent), this
ratio is ${\simeq}0.34$ which is an upper bound for
$F_{L2}(x,Q^2)$  in the
dipole model.\\

In Refs.\cite{Ref18, Ref19, Ref20}, authors show that the ratio of
structure functions in the dipole model is independent of the
dipole cross section  $\sigma_{\mathrm{dip}}$.  Indeed, it is
proportional to the photon-$q\overline{q}$ wave function as
\begin{eqnarray}
F_{L2}(x,Q^2)=g(Q,r,m_{q}) {\leq} \widetilde{g}(z_{m})=0.27139,
\end{eqnarray}
where
\begin{eqnarray}
g(Q,r,m_{q})=\frac{ \int_{0}^{1} dz
|\Psi_{L}(\mathbf{r},z;Q^{2})|^{2}}{ \int_{0}^{1} dz
\bigg{[}|\Psi_{T}(\mathbf{r},z;Q^{2})|^{2}+|\Psi_{L}(\mathbf{r},z;Q^{2})|^{2}\bigg{]}}
\end{eqnarray}
and $m_{q}$ is the mass of the active quark \footnote{For further
discussion see \cite{Ref19}.}. For massless quarks, the function
$g(Q,r,m_{q})$ is defined by the dimensionless variable $z=Qr$ as
the function $\widetilde{g}(z)=g(Q,r,0)$ has a maximum at
$z_{m}=2.5915$ with $\widetilde{g}(z)=0.27139$. It was shown in
literature \cite{Ref21, Ref22, Ref23, Ref24} that, for all
$Q{\geq}0$, $r{\geq}0$ and $m_{q}{\geq}0$, the bound specified by
Eq. (17) for the ratio of structure functions is also valid.\\

\section{Appendix B}

Let us start with the series in  Eq.~(6):
\begin{eqnarray}
 S(v) =
\sum_{k=1}^\infty\frac{k}{(k+1)^2-3(k+1)+4}e^{-(k+1)v}=
\sum_{m=1}^\infty\frac{m-1}{m^2-3m+4}e^{-mv}.
\end{eqnarray}
For large $m$, the terms in the series behave as $e^{-mv}/m$. Noting
that
\begin{eqnarray}
 \sum_{m=1}^\infty \frac{e^{-mv}}{m}=-\ln{(1-e^{-v})}
=-\ln{(1-x)},
\end{eqnarray}
we can subtract this series from Eq.(19) to get a series that converges
as $e^{-mv}/m^2$, so converges even as $v\rightarrow 0$, and add
it back in as $-\ln{(1-e^{-v})}$. This gives
\begin{eqnarray}
S(v)=\sum_{m=1}^\infty\frac{2m-4}{m(m^2-3m+4)}e^{-mv}-\ln{(1-e^{-v})}.
\end{eqnarray}
The result also shows explicitly the divergence for
$x=e^{-v}\rightarrow 1$ or $v\rightarrow 0$.

Let us denote
\begin{eqnarray}
S_{\rm mod}(v)=\sum_{m=1}^\infty\frac{2m-4}{m(m^2-3m+4)}e^{-mv}.
\end{eqnarray}
We can further improve the convergence of the series $S_{\rm mod}(v)$  by subtracting the asymptotic series for $m$
large and adding it back in as the dilogarithm ${\rm Li}_ 2 (e^{-v})=\sum_{m=1}^\infty e^{-mv}/m^2$ .
This gives a series in which the remainder, after $M$ terms, is
of order $1/M^2$
\begin{eqnarray}
S_{\rm mod}(v)=\sum_{m=1}^\infty \bigg{[}\frac{2m-4}{m(m^2-3m+4)}-\frac{2}{m^2}\bigg{]} e^{-mv}+{\rm Li}_ 2 (e^{-v}) .
\end{eqnarray}

In Fig. 6, we show the plots of $S(\upsilon)$ and $S_{\rm
mod}(\upsilon)$, given by Eq. (21) and  Eq. (23), respectively, in
a wide range of $\upsilon$. In both plots, the maximum of $m$ in
the series is chosen to be $M=1000$ with a point wise accuracy
$\sim 1/10^6$. For present purposes value $M\sim 50$ with accuracy
$1/10^4$ or better is sufficient.

\begin{figure}[h]
\includegraphics[width=0.6\textwidth]{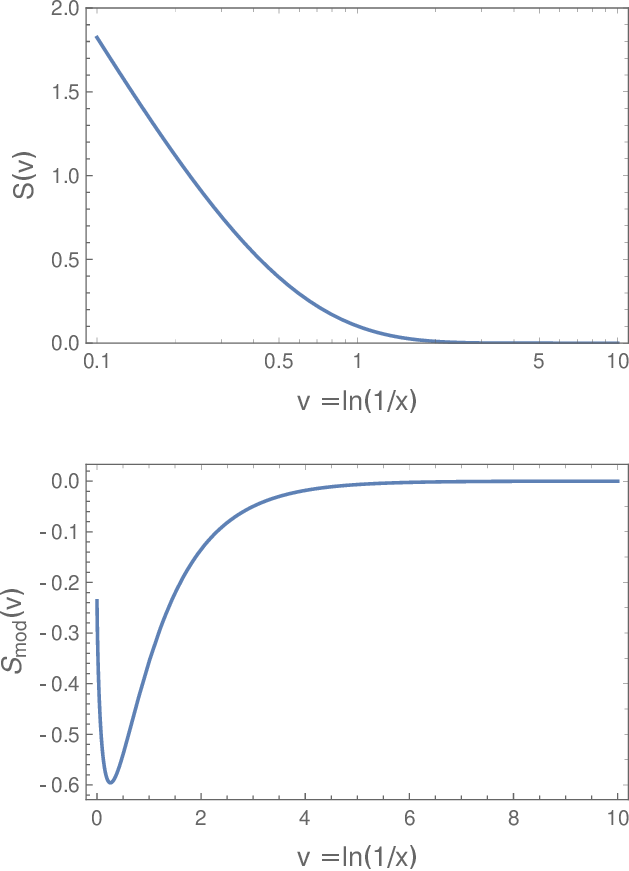}
\caption{ Plots of $S(\upsilon)$ and $S_{\rm mod}(\upsilon)$,
given by Eq. (21) and  Eq. (23), respectively,  in a wide range of
$\upsilon$. Note that $S_{\rm mod}(0) = - 0.2364$. In both plots,
the maximum of $m$ in the series is chosen to be $M=1000$ with a
point wise accuracy $\sim 1/10^6$. }\label{Fig7}
\end{figure}

\begin{table} [h]
\caption{The effective parameters in the BDH expression for
$ F_2(x,Q^2)$ in Eqs. (11) and (12) at small $x$ for
$0.15~\mathrm{GeV}^{2}<Q^{2}<3000~\mathrm{GeV}^{2}$ provided by
the following values. The fixed  parameters are defined by the
Block-Halzen fit to the real photon-proton cross section as
$M^{2}=0.753 \pm 0.068~ \mathrm{GeV}^{2}$, $\mu^2 = 2.82 \pm
0.290~ \mathrm{GeV}^{2}$, and $a_{00}=0.2550\pm 0.016$  \cite{Ref25}.}
\begin{tabular} {cccc}
\toprule \\  \multicolumn{2}{c}{parameters \quad \quad \quad ~~~~~~~~~~~~~~~~value}    \\ &&&\\ \hline \\ &&&\\

  $n$ & \quad  $n=11.49 \pm 0.99$ & &\\

 $\lambda$ & \quad  $2.430~\pm 0.153 $ & &\\

&&&\\

$a_{01}$& \quad  $1.475\times 10^{-1}~\pm 3.025\times10^{-2}$ & &\\

&&&\\

  $a_{10} $  &   \quad  $8.205\times 10^{-4}~~  \pm  4.62\times10^{-4} $  \\

  $a_{11} $  &   \quad   $-5.148\times 10^{-2}\pm 8.19\times10^{-3}$  \\

  $a_{12}$   &    \quad  $-4.725\times 10^{-3}\pm 1.01\times10^{-3}$   \\  &&&\\

 $a_{20}$   &   \quad   $2.217\times 10^{-3}\pm 1.42\times10^{-4} $ \\

 $a_{21}$   &   \quad   $1.244\times 10^{-2}\pm 8.56\times10^{-4}$  \\

 $a_{22}$    &    \quad  $5.958\times 10^{-4}\pm 2.32\times10^{-4} $ \\ &&& \\

\hline

\end{tabular}
\end{table}


\end{document}